\documentclass[reprint,aps,prl,superscriptaddress]{revtex4-2}
\usepackage{graphicx}
\usepackage{dcolumn}
\usepackage{bm}
\usepackage[usenames]{color}
\usepackage{xcolor}
\usepackage{tabularx}
\usepackage{booktabs}
\usepackage{hyperref}
\usepackage{siunitx}
\usepackage[utf8]{inputenc}
\usepackage[english]{babel}
\hypersetup{colorlinks,linkcolor=red,citecolor=blue,urlcolor=blue,final}
\usepackage{amsmath, amsthm, amscd, amssymb} 
\usepackage[mathlines]{lineno}

\newcommand{\Gm}{\Gamma_{\text{m}} } 
\newcommand{\Gfb}{\Gamma_{\text{fb}} }

\begin{document}
\title{Electro-Optical Metalens for Near-Surface Cooling of a Levitated Particle}
\author{Pedro Rosso Gomez}
\affiliation{Nanophotonic Systems Laboratory, Department of Mechanical and Process Engineering, ETH Zurich, 8092 Zurich, Switzerland}
\affiliation{Quantum Center, ETH Zurich, 8083 Zurich, Switzerland}

\author{Bruno Melo}
\affiliation{Nanophotonic Systems Laboratory, Department of Mechanical and Process Engineering, ETH Zurich, 8092 Zurich, Switzerland}
\affiliation{Quantum Center, ETH Zurich, 8083 Zurich, Switzerland}

\author{Adeel Afridi}
\affiliation{Nanophotonic Systems Laboratory, Department of Mechanical and Process Engineering, ETH Zurich, 8092 Zurich, Switzerland}
\affiliation{Quantum Center, ETH Zurich, 8083 Zurich, Switzerland}

\author{Mohamed Sabry Mohamed}
\affiliation{Nanophotonic Systems Laboratory, Department of Mechanical and Process Engineering, ETH Zurich, 8092 Zurich, Switzerland}
\affiliation{Quantum Center, ETH Zurich, 8083 Zurich, Switzerland}

\author{Nadine Meyer}
\affiliation{Nanophotonic Systems Laboratory, Department of Mechanical and Process Engineering, ETH Zurich, 8092 Zurich, Switzerland}
\affiliation{Quantum Center, ETH Zurich, 8083 Zurich, Switzerland}

\author{Romain Quidant}
\affiliation{Nanophotonic Systems Laboratory, Department of Mechanical and Process Engineering, ETH Zurich, 8092 Zurich, Switzerland}
\affiliation{Quantum Center, ETH Zurich, 8083 Zurich, Switzerland}

\date{\today}
\begin{abstract} 
 Advanced levitation optomechanics in the quantum regime requires sophisticated quantum-engineered protocols operating at shorter length and time scales, demanding an increasing control over both optical and electrical potentials. Integrated chip-scale platforms offer a natural route toward such control. Here, we demonstrate optical levitation in vacuum using a miniaturized hybrid electro-optical metalens. Motion control is achieved using a high numerical aperture metalens combined with electrical active feedback using planar electrodes, enabling cooling to low phonon occupations.
 Reheating measurements enable us to rule out any significant influence from electric field noise; an encouraging result for the future development of on-chip levitation platforms operating close to surfaces. This compact integrated platform facilitates the precise manipulation and control of levitated nanoparticles while laying the ground toward exploiting the full capacity of meta-optics to engineer complex optical potentials, marking a significant step forward in technologies for quantum levitation optomechanics. 
 
\end{abstract}

\maketitle

Gaining through miniaturization precise and versatile control over electrons and photons at increasingly short length and time scales has been a key driver of technological innovation, as demonstrated by the development of lasers~\cite{zhou2015chip} or transistors~\cite{cao2023future}. Similarly for fundamental research, this drive for finer control significantly advanced quantum science~\cite{reichel2011atom} and technology~\cite{monroe2013scaling}. As levitation optomechanics progresses into the quantum regime~\cite{delic2020cooling,magrini2021real,tebbenjohanns2021quantum,piotrowski2022simultaneous,dania2025high,Troyer2026,ranfagni2022two,kamba2023revealing}, advanced quantum protocols demand increasingly complicated potential landscapes~\cite{gonzalez2021levitodynamics}, both optical and electrical.  Recent efforts to miniaturize levitation setups from large, table-top experiments to compact, chip-sized modules have garnered significant interest~\cite{shen2021chip,hsu2022single, melo2024vacuum,pi2025levitation, alavi2025compact,liu2026cooling,huang2023metasurface,holman2026trapping}. These novel platforms enable the creation of complex and stable potential landscapes in the proximity of levitated particles, offering future benefits for both fundamental research, such as testing quantum mechanics at large mass and length scales~\cite{RodaLlordes2024,neumeier2024fast,Pino2018,Bose2017,bateman2014near}, and applied research, including force~\cite{hebestreit2018sensing,ranjit2016zeptonewton} and inertial sensing~\cite{monteiro2017optical,kamba2026levitated,melo2024vacuum,Ahn2020,zeng2024optically} for probing non-Newtonian forces~\cite{moore2021searching,Blakemore2021} and applications in seismology~\cite{fuchs2026detection}. Furthermore, many-body physics with macroscopic particles heavily depends on the progress of scalable on-chip platforms in the future~\cite{Siegel2025Optical,holman2026trapping}. \\
Miniaturization strategies for levitated-particle experiments are rapidly developing on multiple fronts. To engineer electrical potentials, approaches span from planar electrode architectures~\cite{DeVoe1992Planar,Kim2010Surface,Alda2016Trapping},  to highly compact 3D architectures~\cite{Brucke2026A,Chou2010Frequency}. These platforms support on-chip atomic shuttling~\cite{Hansel2001Magnetic}, dynamically reconfigurable potentials~\cite{Bohi2009Coherent}, and spatially segmented experimental zones~\cite{Kielpinski2002Architecture}. Recent experiments with levitated nanoparticles have already implemented miniaturized Paul traps~\cite{tomassi2026accelerated,Bonvin2024Hybrid} for dark trapping, as well as planar electrodes for active feedback cooling~\cite{tomassi2026accelerated,melo2024vacuum}.\\
In parallel, optical interfaces fall into three main categories. First, tapered or untapered optical fibers and waveguides are used to deliver light directly onto the chip. Although offering excellent modal confinement, they are constrained to TEM modes and may involve complex nanofabrication~\cite{snyder1983optical,birks1992shape}. Second, molded or 3D-printed micro-optics provide high-NA and compact footprints with comparatively simple fabrication~\cite{juodkazis2016printed}, albeit at the cost of degraded beam quality. 
Third, metasurfaces which offer full control over amplitude, phase, and polarization at the subwavelength scale~\cite{yu2014flat}, enable light focusing~\cite{khorasaninejad2016metalenses}, beam steering~\cite{yu2011light}, and engineering of arbitrary wave fronts ~\cite{ni2013metasurface}. While this is currently the most technologically demanding and typically less efficient than the first two approaches, metasurfaces convince with their unmatched versatility, ultralow footprint thanks to subwavelength thickness and their resilience under high power illumination in ultrahigh vacuum conditions. 
Recent demonstrations include metasurface-based optical manipulation of nanoparticles in liquid~\cite{Tkachenko2018Optical} and vacuum~\cite{shen2021chip}. In vacuum, metasurfaces have recently enabled optical trapping of single nanoparticles~\cite{shen2021chip, sun2024levitation} and even multiple particles~\cite{Siegel2025Optical,sun2024tunable}.\\
Importantly, the proximity to surfaces that enables such fine control over both optical and electric fields also introduces surface-induced noise. This is a well-known limitation in other chip-integrated quantum platforms based on surface-electrode ion traps~\cite{brownnutt2015ion} and could similarly constrain on-chip levitated systems, making the understanding and mitigation of surface-induced noise a central challenge for this miniaturization effort.

In this letter, we introduce an on-chip electro-optical platform that integrates planar electrodes with a high-NA metalens to achieve motion control of a levitated nanoparticle in vacuum down to low phonon occupation. Beyond its trapping performance, we present a comprehensive analysis of the system's reheating sources, quantifying the contributions from residual gas~\cite{gieseler2012subkelvin}, photon recoil heating~\cite{jain2016direct}, electrical surface noise~\cite{teller2021heating} and pointing instabilities~\cite{Savard1997Laser}. We anticipate hybrid metasurfaces will become instrumental to the miniaturization of levitation platforms, promising dynamical and spatial control of potential landscapes inaccessible in standard levitation experiments.  \\

\begin{figure}
\centering
\includegraphics[width=0.48\textwidth]{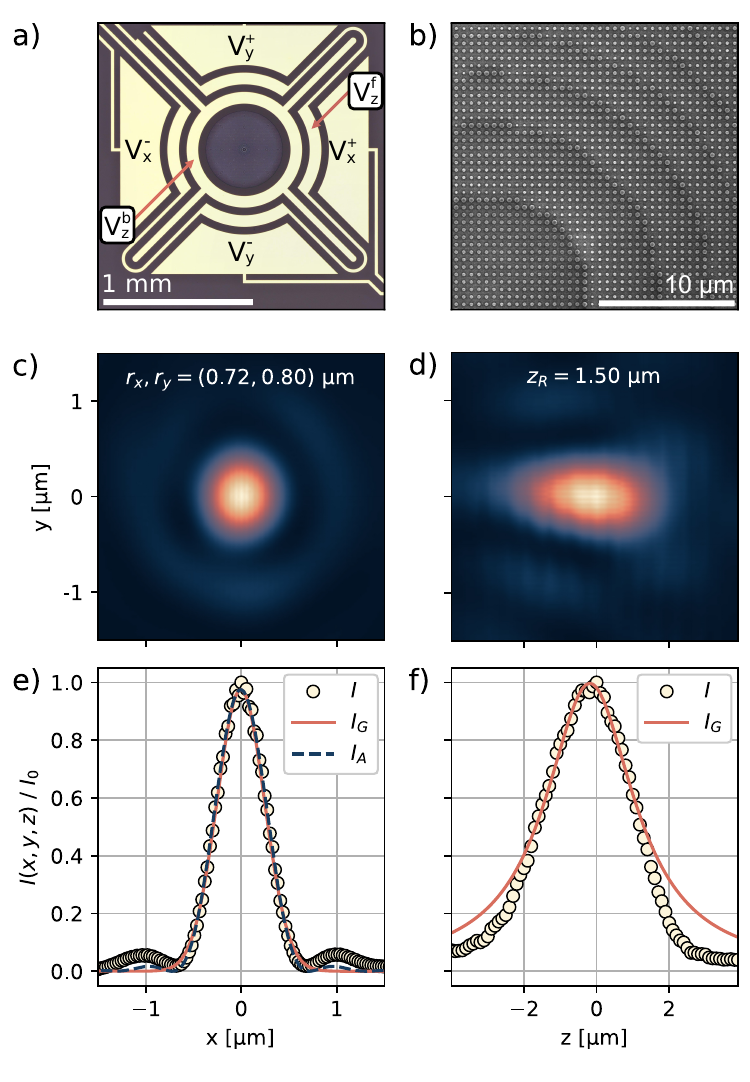}
\caption{\textbf{Hybrid electro-optical metalens:} \textbf{a)} Planar electrode layout (yellow) with central metalens of radius $R_L=\SI{250}{\micro\meter}$ (gray). Feedback voltages $V^\pm_{x,y}$ and $V_z^{f,b}$ are applied to gold electrodes to generate time varying homogeneous electric fields for active cold damping $\Gamma^j_\text{fb}\propto v_j(t)$. \textbf{b)} SEM picture of part of the meta-atoms forming the metalens. Different radii of the meta-atom gives rise to varying local phase shifts $\phi_p(x,y)$ following Eq.~\ref{eq:phase}. \textbf{c)} 2D intensity profile for $y$-polarized light in the $xy$-plane at $z=f$ with airy disc radius $r_x= \SI{0.72}{\micro\meter}$, $r_y = \SI{0.8}{\micro\meter}$,  and \textbf{d)} in the $yz$-plane at $x=0$ with Rayleigh range $z_R=\SI{1.5}{\micro\meter}$. \textbf{e)} Normalized 1D intensity profiles in the focus along $x$ and \textbf{(f)} along $z$. We fit the airy intensity pattern $I_A$ (dashed line) the Gaussian intensity distribution $I_G$ (solid line) to the data (circles). The focusing efficiency equals $\mathrm{FE}=0.67$ and the transmission $T_\mathrm{tot}=0.584$.} 
\label{fig:1}
\end{figure}
The electro-optical metalens chip, shown in Fig.~\ref{fig:1}a consists of two main parts: a high numerical aperture (NA) metalens for optical trapping (central gray disk) and a surrounding planar electrode layout used for center-of-mass cooling (CoM) of the trap particle  (yellow). 
The metalens consists of individual subwavelength pillars as shown in Fig.~\ref{fig:1}b that act as local phase shifters. In order to generate a lens that focuses light of wavelength $\lambda$ along the propagation axis $z$ 
at focal length $f$, we need to apply a spatially varying phase shift to the incident light $\phi_p(x,y)$ that follows the hyperboloidal phase function~\cite{Aieta2012}
\begin{equation}\label{eq:phase}
    \phi_p(x,y) = \frac{2\pi}{\lambda} \left(\sqrt{x^2 +y^2 +f^2} -f\right).
\end{equation}
The metalens is formed by arranging individual silicon meta-atoms, designed to satisfy the phase profile of Eq.~\ref{eq:phase}, in the $xy$-plane in an equidistant grid within the metalens radius $R_L =\SI{250}{\micro\meter}$.\\ 
The metalens focuses the light to a diffraction limited spot forming an Airy pattern. Three-dimensional imaging of the intensity profile is performed to estimate the beam parameters (see SI). Figure~\ref{fig:1}c maps the intensity distribution in the focal plane at $z=f = (125 \pm 5) \SI{}{\micro\meter}$ and Fig.~\ref{fig:1}d at $x=0$. From a 2D fit of the airy disk radius $r_{x,y} = [0.72, 0.80] \pm 0.03 ~\SI{}{\mu \meter}$, we estimate the NA of the metalens to 
\(
    \text{NA}_{x,y} = 1.22\: \lambda/r_{x,y} = [0.81, 0.90],
\)
in good agreement with the simulated design $\text{NA}=0.9$. The beam waists of the corresponding Gaussian beam then result in $w_x = (0.37 \pm 0.03) ~\SI{}{\micro\meter}$ and $w_y = (0.42 \pm 0.03)~\SI{}{\micro\meter}$. The focus shows spatial asymmetries ($w_x < w_y$) due to strong focusing~\cite{novotny2012principles}.\\
The 1D intensity distribution $I(x)$ with $y=0,z=f$ follows well the Gaussian beam profile (red solid line) the Airy function (blue dashed) close to the focus as depicted in Fig.~\ref{fig:1}e. The Rayleigh range is estimated from a Gaussian beam profile (red solid line) to  $z_R=(1.50 \pm 0.10)\SI{}{\mu\meter}$ using the 1D intensity distribution $I(z)$ with $x=y=0$ as displayed in Fig.~\ref{fig:1}f.
Additionally to the beam shape, we characterize the metalens efficiency. The transmission of the metalens is given by $T_\mathrm{tot} = P_t/P_0= 0.584$ where $P_t$ is the total transmitted power and $P_0$ the estimated power in front of the lens aperture. Furthermore, we define the focusing efficiency as $\text{FE} =\int_0^{r} P(r) r \: dr \:/P_t $ where $r=1.5 \times\text{FWHM}$.
We estimate $\text{FE} = 0.67 $ due to diffraction and unfocused light passing through the metalens.
For more details on the meta-atom library and fabrication see SI.\\

\begin{figure}
\centering
\includegraphics[width=0.48\textwidth]{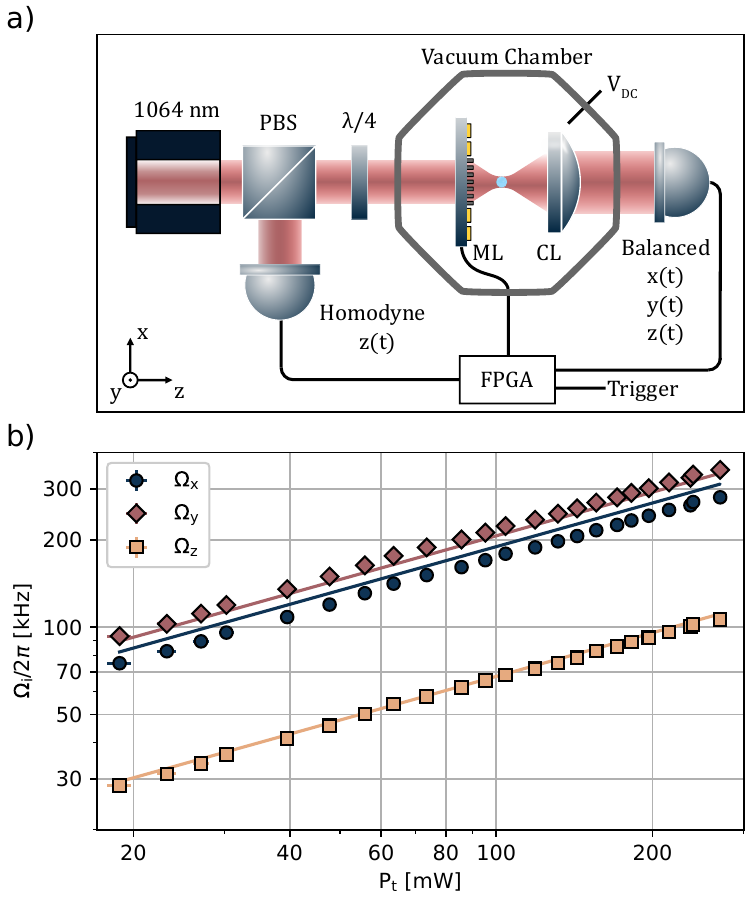}
\caption{\textbf{Particle levitation using a hybrid metalens:} 
\textbf{a)} Experimental setup using a hybrid metalens to focus a laser ($\lambda=\SI{1064}{\nano\meter}$, $P_0$). A particle (blue) is trapped at the focus and the scattered light is collected with a collection lens (CL) in forward direction for balanced split detection of $x(t),y(t),z(t)$ and through the metalens for homodyne detection of $z(t)$. Quarter waveplates ($\lambda/4$) and  polarizing beam splitter (PBS) are used for polarisation control. Electrical feedback is applied with an FPGA to the planar gold electrodes (yellow). High voltage $ V_\text{DC}$ allows to control the charge $n_q$.  
\textbf{b)} Mechanical eigenfrequencies following the expected dependence on the optical power $\Omega_j \propto \sqrt{P_t}$ (see SI). The data follows well the theory (solid lines) where $z_R$ is estimated from $\Omega_j$ and $w_{x,y}$. }
\label{fig:2}
\end{figure}
The tight laser focus from the metalens (wavelength $\lambda= 1064$ nm, optical power $P_0$) enables optical trapping of a single charged silica nanosphere of radius $R$ at the intensity maximum as displayed in the experimental setup shown in Fig.~\ref{fig:2}a. The particle's CoM motion is detected by collecting the scattered light in the forward direction using balanced split detection for $x(t)$, $y(t)$, $z(t)$~\cite{gieseler_thesis}, and in the backward direction using homodyne detection through the metalens for $z(t)$~\cite{tebbenjohanns2019optimal}, where $z$ denotes the optical axis.
The particle's charge $q = n_q \times e^{-}$ is controlled with a single elementary charge precision by ionizing gas molecules with a high DC voltage $V_\text{DC}$ at moderate vacuum~\cite{frimmer2017controlling}. 
Remarkably, the metalens shows no detectable sign of short-term degradation under constant high power illumination in vacuum and repeated unshielded particle nebulization. \\
%
The optical harmonic potential generated by the metalens gives rise to well defined mechanical eigenmodes 
that follow the well known power dependence of $\Omega_j\propto \sqrt{P_t}$~\cite{gonzalez2019theory} with $j =x,y,z$ as depicted in Fig.~\ref{fig:2}b. At maximum power $P_t =\SI{270}{\milli\watt} $, we reach $(\Omega_x,\Omega_y,\Omega_z)/2\pi \simeq[280,348,106]~\text{kHz}$, matching standard microscope objectives with the same NA~\cite{gieseler2012subkelvin}. 
To estimate the nonlinearities in the metalens potential, we reverse the Gibbs probability density function~\cite{conangla2020extending} to reconstruct the optical potential $\Psi(x,y,z)$ seen by the particle at $p_\text{gas} \gtrsim \SI{16}{\milli\bar}$
\begin{equation}
    \rho_\infty(x,y,z)= \frac{1}{Z} \exp\left(-\frac{\Psi(x,y,z)}{k_B T}\right)
\end{equation}
where $\rho_\infty(x,y,z)$ is the probability distribution of the position timetrace $x(t), y(t), z(t)$, $Z$ a normalization factor, $T=\SI{300}{\kelvin}$ room temperature and $k_B$ the Boltzmann constant. The reconstructed potential matches well the harmonic approximation with a Duffing nonlinearity $\xi_x,\xi_y,\xi_{z,f},\xi_{z,b} = [0.05, -61, 29, 38] \SI{}{\micro\meter^2}$~\cite{cuairan2022precision} (see SI).\\

\begin{figure*}
\centering
\includegraphics[width=\textwidth]{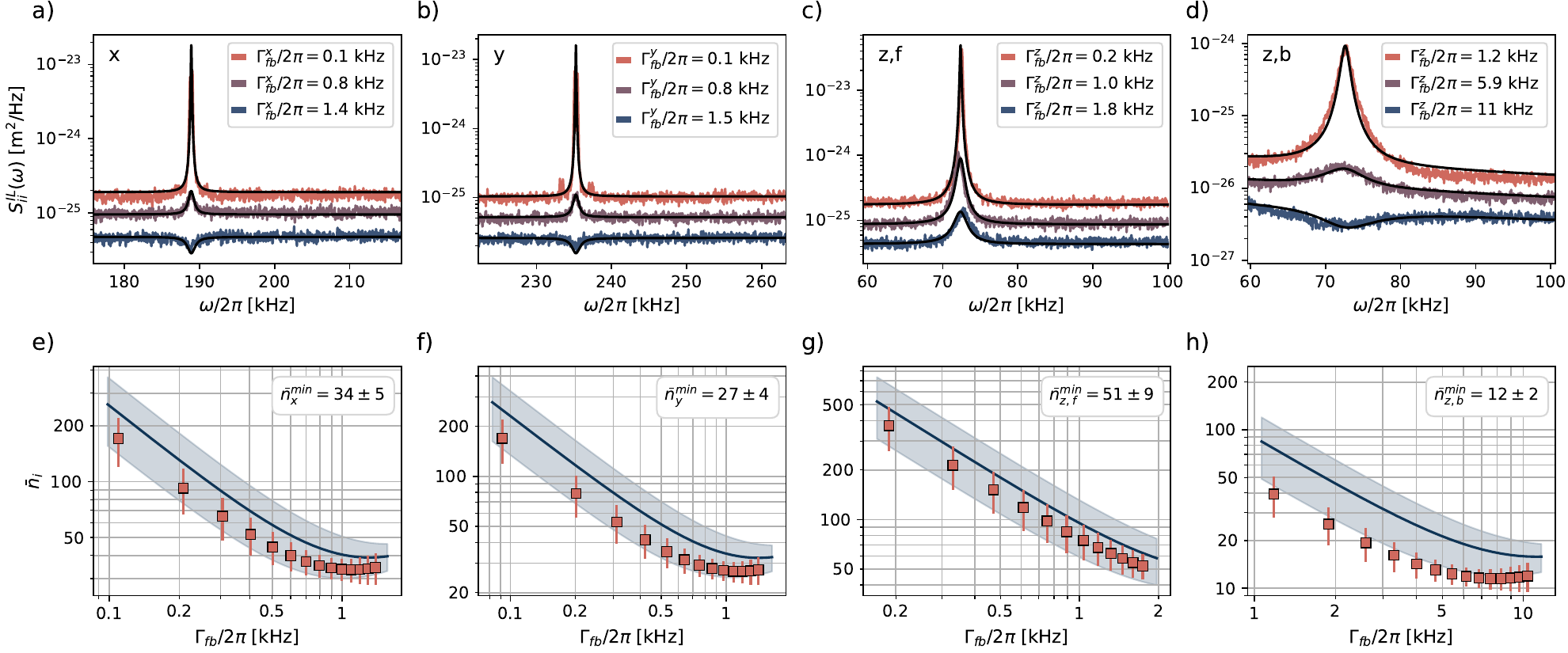}
\caption{\textbf{CoM cooling with hybrid metalens:} PSD of the displacement $S_{jj}^\text{IL}(\omega)$ along \textbf{a)} $x(t)$,  \textbf{b)} $y(t)$ , \textbf{c)} $z(t)$ forward, \textbf{d)} $z(t)$ backward for different feedback strengths $\Gamma_\text{fb}^j$ measured with an IL detector. Noise squashing is observed on nearly all detectors. The data is fitted to Eq.~S58 (solid line, see SI) and allows the extraction of $\Gamma^j_\text{fb}, \Gamma_m, \Omega_j, \sigma_j$ and $\sigma^j_\xi$. The corresponding phonon occupation \textbf{e)} $\bar{n}_x$,  \textbf{f)} $\bar{n}_y$ , \textbf{g)} $\bar{n}_z$ forward, \textbf{h)} $\bar{n}_z$ backward is estimated with Eq.~\ref{eq:n_bar_j} for varying $\Gamma^j_\text{fb}$. The minimal phonon occupations equal $[\bar{n}^\text{min}_x,\bar{n}^\text{min}_y,\bar{n}^\text{min}_{z,f},\bar{n}^\text{min}_{z,b}] = [34, 27, 51, 12]$ for a particle with $R=\SI{109}{\nano\meter}$ and $n_q=13$,  at $p_{gas} = (1.5 \pm 0.15) \cdot 10^{-7} \SI{}{\milli \bar}$ agree well with theory (solid line) based on independent estimations of $m, \Gamma_m, \Omega_j, \sigma_j$ and $\sigma^j_\xi$. Error bars are dominated by calibration uncertainties and the shaded area accounts for calibration, pressure and waist uncertainties.}
\label{fig:3}
\end{figure*}

Stabilization and cooling of the CoM motion in vacuum is implemented by electrical measurement-based feedback~\cite{conangla2018motion, tebbenjohanns2019cold,melo2024vacuum}. The particle's motional signal is delayed by a quarter of a period using an FPGA (see Fig.~\ref{fig:2}a).  This electrical feedback signal is applied to the planar electrodes around the metalens depicted in Fig.~\ref{fig:1}a. 
Two pairs of opposing electrodes $V_{x,y}^{\pm}$ and two independent central rings $V_{z}^{f,b}$ generate independent homogeneous electrical fields $E_x(t),E_y(t), E_z(t)$~\cite{melo2024vacuum}. The feedback signals give rise to a linear damping force $\vec{F}_\text{fb}(t)\propto \vec{\Gamma}_\text{fb}\cdot\vec{v}(t)$ where $\vec{v}(t)$ is the particle's velocity and $\vec{\Gamma}_\text{fb} =[\Gfb^x,\Gfb^y,\Gfb^z]$ the electrcial feedback damping. The two ring electrodes ensure independent feedback along $z$ with stable forward (outer ring) and sensitive backward detection (inner ring) to enable efficient cooling without particle loss. Note that, only in-loop detection (IL) is used~\cite{conangla2019optimal,tebbenjohanns2019cold,Melo2025}. \\
%
The average phonon occupation  $\bar{n}_j$ of each eigenmode is used to assess the cooling performance (see SI)
 \begin{equation}\label{eq:n_bar_j}
      \bar{n}_j = \frac{k_B T^j_\text{eff}}{\hbar \Omega_j} =\frac{m\Omega_j}{2\hbar }   \left(\frac{\sigma_j^2 }{ m^2\Omega_j^2 (\Gamma_m+\Gamma^j_\text{fb})}    + \frac{(\Gamma^j_\text{fb}\sigma^j_\xi)^2 }{ (\Gamma_m+\Gamma^j_\text{fb})}   \right)
 \end{equation}
where  $\sigma_j^2  =\sigma^2_{m} + \sigma^2_{\text{ph},j}+ \sigma^2_{v,j}$ is the force noise variance,  with $\Gm,\sigma_{m} \propto p_\text{gas}$ being related to the residual gas pressure~\cite{gieseler2012subkelvin}, $\sigma_{\text{ph},j}$ to the photon recoil~\cite{jain2016direct}, and $\sigma_{v,j}$ to the electrical noise~\cite{teller2021heating}. For 
$\Gamma_m \gg \Gfb^j$ one recovers the well-known expression $\Gamma_m T = T^j_\text{eff}(\Gamma_m + \Gfb^j)$.\\ 
The experimental power spectral density (PSD) of the measured particle displacement $S_\text{jj}^\text{IL}(\omega)$ is fitted to the theoretical expression Eq.~S43 (see SI) to extract $\Omega_j, \sigma_j,\Gfb^j, \Gamma_m$ and $\sigma_\xi^j$. From the calibration at elevated pressures, we estimate $m$ (see SI).  
For increasing $\Gamma^j_\text{fb}$, the amplitude of the PSD of the particle displacement $S^\text{IL}_\text{jj}(\Omega_j)$ (see Fig.~\ref{fig:3}a-d) and the phonon occupation $\bar{n}_j$ along $x,y,z$ forward, and $z$ backward (see Fig.~\ref{fig:3}e-h) decrease, as expected from Eq.~\ref{eq:n_bar_j}. For a particle with $R=\SI{109}{\nano\meter}$ and $n_q=13$, the minimum phonon occupation saturates at $\bar{n}_j^\text{min} = [34, 27, 12]$ at optimal $\Gfb^j$ at constant pressure $p_{gas} = (1.5 \pm 0.15) \cdot 10^{-7} \SI{}{\milli \bar}$ in good agreement with the theory predictions $\bar{n}_j^\text{th} = [39, 32, 16]$ from Eq.~\ref{eq:n_bar_j} (solid lines) where the proportionality factor between the electrical signal gain $k_d$ and $\Gamma^j_\mathrm{fb} = a  k_d$ is the only fit parameter. The discrepancy between theory and experimental values arises from the overestimation of the photon recoil force $\sigma^2_{ph,j} \propto m^2$, dominated by the mass uncertainty.\\

\begin{figure}[h]
\centering
\includegraphics[width=0.47\textwidth]{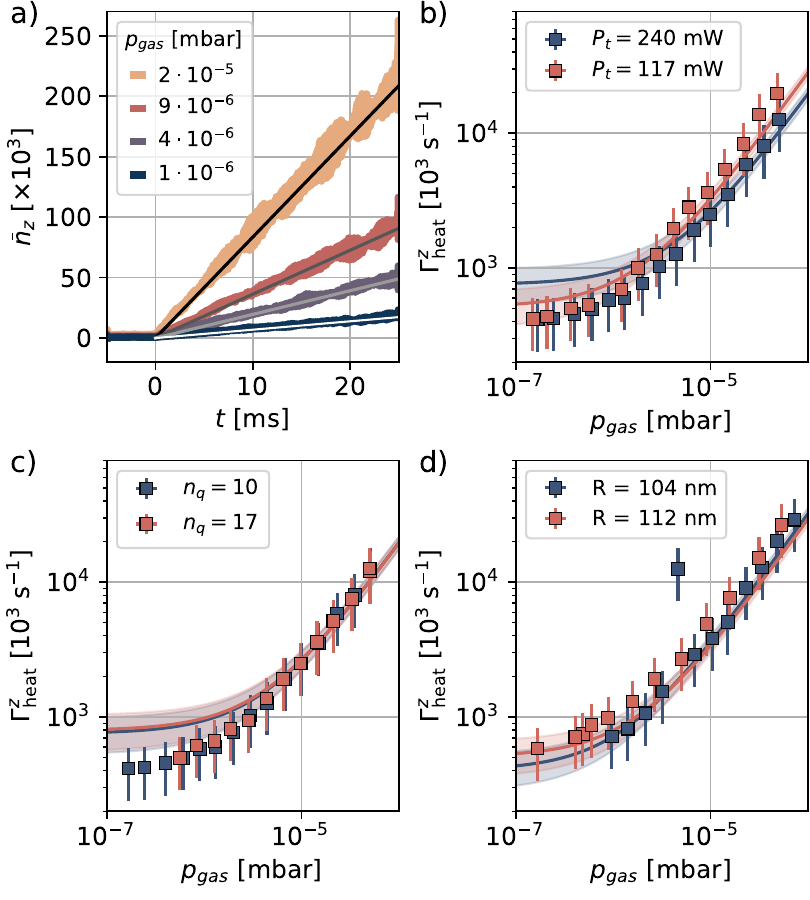}
\caption{\textbf{Reheating of CoM motion:}
\textbf{a)} Phonon occupation $\bar{n}_z(t)$ with $\Gamma_\text{fb}^z>0$ for $t\leq 0$ and  $\Gamma_\text{fb}^z=0$ for $t>0$ at various pressure $p_\text{gas}$ based on 300 averaged trajectories. The linear fit (solid lines) estimates the heating rate $\Gamma^j_\text{heat}$.
$\Gamma^j_\text{heat}$ versus $p_\text{gas}$ for \textbf{b)} different trapping powers $P_t$, \textbf{c)} different charges $n_q$, and \textbf{d)} different particle sizes $R$. The errorbars correspond to the standard deviation of the 300 independent trajectories. The solid lines are theoretical estimations based on independent measurements of $p_\text{gas}, R, n_q, P_t, w_x,w_y$ and $z_R$ and the shaded areas represent one standard deviation of those parameters. The parameters are stated in the SI. 
}
\label{fig:4}
\end{figure}
To identify the noise sources limiting our lowest phonon occupation $\bar{n}_j^\text{min}$, we perform reheating measurements of the pre-cooled CoM motion. Starting from a low phonon state $\bar{n}_j \ll \bar{n}_\mathrm{th}= k_B T/(\hbar \Omega_j)$, we switch off the feedback ($\Gamma_\text{fb}=0$) at time $t=0$ and allow the particle to evolve in the optical trap at different pressures $p_\text{gas}$ before switching the feedback back on ($\Gamma_\text{fb}^j\neq 0$)  at $t=25 ~\SI{}{\milli \second}$. We average over 300  trajectories~\cite{gieseler2013thermal} as depicted in Fig.~\ref{fig:4}a. The phonon gain is linear for short times following $\bar{n}_j(t) = \bar{n}^0_j + \Gamma_\text{heat}^j t$~\cite{jain2016direct} where $\bar{n}_j^0$ is the initial phonon occupation at $t=0$. The total phonon heating rate $\Gamma_\text{heat}^j$ is estimated from fitting $\bar{n}_j(t)$ at various pressures $p_\text{gas}$ (solid lines) and  is given by $\Gamma_\text{heat}^j = \sigma_j^2/(2m\hbar\Omega_j) = \Gamma^{(p)}_j + \Gamma^{(r)}_j + \Gamma^{(e)}_j$. The primary heating sources are residual gas molecules $\Gamma^{(p)}_j = \Gm k_B T /(\hbar \Omega_j)$ and photon recoil heating $\Gamma^{(r)}_j~\propto P_t\lambda^3$~\cite{gonzalez2019theory,jain2016direct}, which can be mitigated through improved vacuum conditions and the use of dark/hybrid trapping potentials~\cite{tomassi2026accelerated,Bonvin2024Hybrid}, respectively. Additionally, the short focal length of the metalens and the particle's proximity to its surface can induce electrical field noise heating $\Gamma^{(e)}_j$, driven by voltage noise from the feedback electronics and dielectric surface noise analogous to ion traps~\cite{teller2021heating,brownnutt2015ion}. 
To estimate the influence of the different heating sources,reheating measurements are performed at different optical powers $P_t$ (see Fig.\ref{fig:4}b), electrical charges $n_q$ (see Fig.\ref{fig:4}c) and particle sizes $R$ (see Fig.\ref{fig:4}d). Other experimental parameters can be found in Tab.~S1-S4 (see SI) and were kept as stable as possible throughout the measurements. 
The solid lines represent the theoretical prediction of $\Gamma_\text{heat}^j$ (see SI) with the shaded areas accounting for calibration errors in phonon 
estimation and fitting. From Fig~\ref{fig:4}b-d we conclude that photon recoil heating $\Gamma^{(r)}_j$ is the main heating source at low pressures $p_{gas} < 10^{-6}\SI{}{\milli \bar}$ (see Fig.~\ref{fig:4}b) and that $\Gamma^{(e)}_j$ is negligible in comparison to  $\Gamma^{(p)}_j $ and $ \Gamma^{(r)}_j$ (see Fig.~\ref{fig:4}c) in the parameter range used here. We estimate the lowest heating rate $\Gamma_\text{heat}^j =[86.71, 54.73, 419.72] \times 10^3$ phonons per second at $p_{gas} = (1.5\pm 0.15) \cdot 10^{-7} \SI{}{\milli \bar}$ and $P_t =\SI{117}{\milli\watt}$, slightly below the expected theoretical values $\Gamma^{(p)}_j =[27.65, 29.53, 32.04]\times 10^3 \SI{}{\second^{-1}}$  and $\Gamma^{(r)}_j =[140.74, 128.81, 483.18]\times 10^3\SI{}{\second^{-1}}$ (see SI). For the electrical heating, we estimate theoretically  $\Gamma^{(e)}_j = [1.26, 0.82, 28.84] \times 10^3$ phonons per second, $S_{EE}^j(\Omega)=[1.3, 1.2, 14]\cdot 10^{-6}$\SI{}{\volt^2 \meter^{-2} \hertz^{-1}} assuming $n_q = 13$, $f=125~\SI{}{\micro\meter}$, $m=\SI{11.8}{\femto \gram}$ and $\Omega_j / 2 \pi = [185, 230, 70]~\SI{}{\kilo \hertz}$ (see SI). The reheating rate induced by dielectric loss in the material~\cite{teller2021heating} to be $\Gamma_z^{(d)} = 27$ phonons per second (see SI).

\paragraph{Discussion and outlook}
In summary, we have demonstrated control of the center-of-mass motion of a levitated nanoparticle above a hybrid electro-optical chip that integrates miniaturized optical and electrical potentials. The performance of both trapping and cooling is comparable to that achieved in conventional experimental platforms involving bulk optical elements and electrodes. While this prototype already combines the essential functionalities required for nanoparticle trapping and CoM cooling, the broader potential of planar electrodes and metasurfaces remains largely unexplored. In particular, their flexibility could enable dynamic potential control, confinement on smaller scales, engineered higher-order potentials, and access to many-body physics, providing a promising route toward advanced protocols for quantum control and quantum engineering. The miniaturized architecture is also particularly attractive for space-constrained environments, including cryogenic platforms, compact sensing applications, and space-based experiments.\\
In addition, we have investigated the relevant heating mechanisms and find that they are not more detrimental than in established experiments. In particular, electrical noise associated with nearby surfaces does not appear to impose a significant limitation on the demonstrated performance. Beyond its relevance for integrated nanoparticle control, this platform therefore 
offers important insight for future experiments involving levitated particles in close proximity to surfaces.

\vspace{0.5cm}
\textbf{Acknowledgments}
\noindent We acknowledge valuable discussions with the QXtreme consortium. 
This research was supported by the European Research Council (ERC) under the grant Agreement No. [951234] (Q-Xtreme ERC-2020-SyG).

\textbf{Author contributions} - 
MSM designed and implemented the optical setup. PRG fabricated the device, PRG and BM performed the measurements and analyzed the data.   %
PRG and AA performed the numerical simulations,
 NM and RQ conceptualized the experiments. All authors discussed the results and contributed to writing the manuscript.\\

\textbf{Competing Interests} - 
The authors declare no competing interests.\\

\textbf{Data availability} - 
The data supporting the findings of this study are available within the
article and its Supplementary Information. The source data files are
available via the ETH Zürich Research Collection at TBA.


\newpage
%

\end{document}